# Unraveling the magnetic structure of YbNiSn single crystal via crystal growth and neutron diffraction

H. C. Wu[1, *], A. Nakamura[2], D. Okuyama[1], K. Nawa[1], D. Aoki[2], and T. J. Sato[1]

[1]*Institute of Multidisciplinary Research for Advanced Materials, Tohoku University, Sendai 980-8577, Japan*

[2]*Institute for Materials Research, Tohoku University, Oarai, Ibaraki 311-1313, Japan*

(Dated: May 17, 2023)

*wu.hung.cheng.d2@tohoku.ac.jp

## Abstract

Neutron and x-ray diffraction experiments were performed on the ternary intermetallic compound YbNiSn, formerly categorized as a ferromagnetic Kondo compound. At zero field, an increase in scattering intensity was observed on top of allowed and forbidden nuclear reflections below $T_c$, breaking the reflection condition of the crystal symmetry *Pnma*. This indicates that the magnetic structure of YbNiSn is antiferromagnetic-type, rather than the previously proposed simple collinear ferromagnetic structure. Temperature dependence of the scattering intensity of the 011 reflection confirmed the magnetic ordering at 5.77(2) K. No incommensurate satellite reflection was observed at 2.5 K. By applying external magnetic field of 1 T along the *a* axis, the magnetic intensity at the nuclear-forbidden 001 position was suppressed, while a slight enhancement at the nuclear-allowed 002 position was observed. This suggests a spin-flip transition under the external magnetic field along the *a* axis in YbNiSn. The proposed magnetic structures at zero field and 1 T correspond to the magnetic space groups of *Pn'm'a* and *Pnm'a'*, respectively. The piezomagnetic effect and the switch between the two magnetic space groups by the external stress, which could be detected by the anomalous Hall effect, are proposed.

## I. INTRODUCTION

The equiatomic *RE*NiSn (*RE* = Rare earth element) compounds with TiNiSi-type [1-10] structure exhibit diverse physical properties, such as heavy fermion (HF) ferromagnet [1-2], incommensurate sinusoidally modulated structure [4, 8, 9], complex metamagnetism [5-7, 11], and topological kondo insulator [10]. In actinide compounds, the ferromagnetic superconductors, URhGe and UCoGe [12], form this crystal structure. The field-reentrant superconductivity in URhGe is induced by the ferromagnetic fluctuations associated with the spin-reorientation [12], which might be due to this peculiar TiNiSi-type structure. The magnetic transition temperature as a function of the atomic number of *RE* for *RE*NiSn compounds exhibits a non-monotonic change, hinting that the tunable electronic configuration of the different *RE* leads to the alternation of the magnetic interaction and magnetic ground state.

The Doniach phase diagram [13] theoretically represents the competition between inter-site Ruderman-Kittel-Kasuya-Yosida (RKKY) and on-site Kondo interactions. A quantum critical point separated by magnetic order, paramagnetic fermi liquid, and non-fermi liquid is expected when energy scales of the two interactions are comparable. Among the *RE*NiSn family, Ce- [3] and Yb- [1-2] based materials are heavy fermion systems, and supposedly at each side of the quantum critical point. The former was found to be a non-magnetic Kondo semiconductor with the formation of a small energy gap. With chemical substitutions on the Ni atom by electron ($CeNi_{1-x}Cu_xSn$) doping [14] or a larger ($CePt_{1-y}Cu_ySn$) atom [15], the development of a magnetically ordered Kondo-lattice state was seen at $x = 0.13$ and $y = 0.33$ due to the reduction of *c-f* hybridization [14, 15]. The latter compound, YbNiSn, is a hole counterpart of $Ce^{3+}$($4f^1$, one electron) with a metallic behavior. It exhibits a magnetic transition below $T_c$ (critical temperature) of 5.5 K with a large electronic specific heat γ-coefficient of ~300 mJ/mol $K^2$ [1] and undergoes multiple field-induced metastable transitions [1-2]. Kasaya *et al*. [1] reported an antiferromagnetic-like transition when a low-magnetic field (0.5 T) is applied along the *a* axis, indicating that the FM ground state of the spin structure is unlikely. In addition, the electrical resistivity as a function of magnetic field at 2.5 K shows a broad peak across the metamagnetic transition, followed by a dramatic decrease up to 7 T [1]. The Hall effect also reveals a strong correlation between magnetic and electrical properties [16]. In contrast to the intensive studies on the diverse physical properties of this material, the ground state magnetic structure was only incompletely studied, and the results are still contradicting. From the bulk magnetic measurement, a canted antiferromagnetic (C-AFM) structure was proposed [1], where two antiparallel sublattices couples along the [100] direction with a sight canting of magnetic moment towards the [001] direction. This results in the weak spontaneous magnetization moment along the *c* axis. ($m_a > m_c$). On the other hand, much microscopic technique, a neutron diffraction, proposes a simple collinear ferromagnetic structure where the moments direct to the [001] direction [2].

The ambiguous magnetic structures [1, 2] motivate us revisiting the zero-field magnetic structure in YbNiSn. In this work, neutron and x-ray diffraction experiments were performed using the high-quality Bridgman-grown single crystal to verify the magnetic structure. The observation of magnetic intensity appearing both at the allowed and forbidden nuclear reflection positions at zero field below $T_c$ indicates the existence of antiferromagnetic-type ordering in YbNiSn. Under external field of 1 T applied along the *a* axis, the spin-flip of a half of the spins is suggested from the recovery of the reflection condition for the 00*l* reflections, $l = 2n$. We speculate that the spin flip is the origin of metamagnetic transition in YbNiSn.

## II. EXPERIMENTAL METHODS

High-quality single crystals of YbNiSn with the residual resistivity ratio (RRR) of 55-80 were grown using the Bridgman method. Single crystals were oriented by the Laue photograph, and were cut using a spark cutter. The single crystal was characterized using a four-circle X-ray diffractometer (Rigaku RA MultiMax-9 with Huber Eulerian cradle) with the Mo $K_\alpha$ radiation equipped with a pyrolytic graphite (PG) monochromator. The acceleration voltage of 30 kV was used to suppress the higher harmonic x-ray, and the current was set to 30 mA. A selected single crystal was further formed into a spherical shape with a diameter of 50 μm, and used in the detail x-ray diffraction measurement. For this measurement, $\theta$-$2\theta$ scans were performed to collect the integrated intensity to obtain the $|F|^2$ table, where the Lorentz and polarization factor is corrected. The ShelXle software with the full-matrix least-squares method was used for the crystal structure refinement [17].

Magnetization measurements at 2.0 K were performed using the magnetic property measurement system (Quantum Design, MPMS). A single crystal of ~1.5 x 1.5 x 0.2 mm$^3$ was used in this measurement, and the crystal was aligned with either *H* // *a*, *b*, or the *c* axis.

Neutron diffraction experiments were performed using the GPTAS (4G) triple-axis thermal neutron spectrometer installed at JRR-3, Tokai, Japan, with a vertically focusing PG 002 monochromator. The spectrometer was used in the two-axis mode without the analyzer. The incident neutron energy ($E_i$) was selected to be 30.5 meV (1.6377 Å). The wavelength was calibrated using standard $Al_2O_3$ before performing the neutron diffraction experiment. To eliminate the higher harmonic neutron contribution, PG filter was inserted at the upstream of the PG monochromator. The single crystal shown in Fig. 2 (a) was aligned with its 0*kl* plane coinciding to the horizontal scattering plane. The sample was sealed in a standard aluminum sample can with $^4$He exchange gas. A $^4$He closed cycle refrigerator, a Gifford-McMahon (GM) cooler, was used in the neutron diffraction experiment with a base temperature of 2.5 K. An electromagnet (dice-shaped magnet) was used to generate vertical magnetic field up to 1 T.

## III. RESULTS AND DISCUSSION

### A. Crystal structure determination and isothermal magnetization.

First, we revisited the crystal structure of YbNiSn using the single-crystal x-ray diffraction analysis. Condition for the appearance of the Bragg reflections was checked with several crystals, obtained from one batch of Bridgman-grown chunk. The results are fully consistent with the previously reported *Pnma* space group. A spherical crystal with a diameter of 50 μm was then prepared to obtain the complete structure factor table. After the least-square refinement, with the initial fractional coordinates taken from the earlier report [2], we have obtained a reasonable crystal structure model. A comparison between the calculated and observed structure factors is shown in Fig. 1 (a). The refined parameters and reliable factors are summarized in Table. I, in agreement with previous report [2]. Our refinement confirms the non-symmorphic structure with a distorted honeycomb lattice along *b* axis for YbNiSn, as displayed in Fig. 1 (b).

The isothermal *M-H* curves up to 5 T along the *a*, *b*, and *c* axes at 2.0 K are presented in Fig. 1 (c). The linear *M-H* curve is observed for *H* // *b* axis and the magnetization quickly reaches to saturation at 0.7 μ$_B$ for *H* // *c* axis. On the other hand, when *H* is parallel to the *a* axis, a metamagnetic transition with a jump of magnetization was observed. By taking differential d*M*/d*H* shown in Fig. 1(d), the metamagnetic anomaly field was estimated as $H_{C1}$ = 0.84 T. It is further noted that above $H_{C1}$ the magnetization increases linearly again, with a clear slope change at $H_{C1}$. Finally, the magnetization reaches to the saturation above $H_{C2}$ ~1.6

T. This highly anisotropic behavior in magnetization is consistent with previous reports [1, 2].

## B. Temperature dependence of magnetic intensity at zero field.

The earlier powder neutron diffraction indicated that the most intense magnetic reflection is 011 [2]. Therefore, the 01-1 reflection was first selected to study its temperature dependence. Figures 2 (b, c) show the diffraction patterns taken both with the omega- and $\theta$-$2\theta$ scans around the 01-1 reflections. The increase in scattering intensity between 2.5 K and 7 K verifies the development of magnetic signal. The temperature dependence of the peak intensity measured at the 011 position is shown in Fig. 2 (d). The scattering intensity gradually decreases with increasing temperature and remains constant above 5.77(2) K, indicating that the magnetic ordering temperature in YbNiSn is 5.77(2) K. This ordering temperature is consistent with the earlier magnetization measurement [1].

Next, $\theta$-$2\theta$ scans were performed at temperatures below (2.5 K < $T_c$) and above (7 K > $T_c$) the transition temperature under zero magnetic field. All reachable reflections in the $0kl$ plane were measured, as shown in Fig. 3 (a). As representative examples, results for the four reflections are presented in Fig. 3(c-f). For the 00-2 reflection shown in Fig. 3(d), no change of scattering intensity was seen below and above $T_c$, while the 00-1, 01-1, and 01-2 reflections exhibit intensity increase when going into the ordered phase below $T_c$, as shown in Figs. 3(c), 3(e), and 3(f), respectively. Combined with the results for the other reflections, we conclude that the reflection condition of *Pnma* along the $l$ direction, $00l$ with $l = 2n+1$, is fully relaxed for the magnetic reflections. This rules out the candidate with a collinear ferromagnetic model (Scenario II) [2]. It should be noted that no incommensurate satellite reflection was observed along the higher symmetry $0k0$, $00l$, $0k$-$k$, and $0kk$ directions at 2.5 K; as an example, a scan along $00l$ is shown in Fig. 3(b).

## C. Magnetic structure determination at zero field

Integrated intensity of the reflections was estimated from the results of the $\theta$-$2\theta$ scans. The structure factor $|F|$ table was then obtained by correcting them by the Lorentz factor. Reflections with $I > \sigma$ were used in the magnetic structure analysis below, where $\sigma$ is the uncertainty or standard deviation. The intensity of thirteen nuclear reflections measured at 7 K, with $I > 3\sigma$, was used to obtain the scale factor, as well as the crystal-structure parameters as detailed below. For the magnetic structure analysis, the intensity of nine magnetic reflections measured at 2.5 K was used. It should be noted that finite intensity was detected at the 00-1 position at the paramagnetic temperature 7 K in the neutron diffraction experiment. The 00-1 reflection is forbidden for the *Pnma* space group, which was confirmed in our x-ray study. Hence, the weak 00-1 reflection detected in the neutron experiment could be from extrinsic origins, such as double reflections or higher-harmonic neutrons, or could suggest intrinsic symmetry lowering of the large-sized crystal. In either case, the intensity ratio between the strongest reflection 020 and 00-1 is approximately 0.1%, and hence, the appearance of the 00-1 reflection will not affect the analysis of magnetic structure below.

Magnetic representation analysis was used to identify the possible candidates for the magnetic structure. The reduction of the magnetic representation for the Yb (4$c$) site of the *Pnma* space group suggested eight one-dimensional irreducible (IR) representations (IR1 to IR8). Table II summarizes the basis vectors corresponding to each IR for the four Yb sites, the magnetic category, and corresponding magnetic space group. After considering the reflection condition of each magnetic structure generated from eight IRs, we found that only IR7 is consistent with the observed magnetic reflection positions. This corresponds to the magnetic space group of *Pn'm'a*. The magnetic structure is a superposition of antiferromagnetic *a*- component (AFM$_{\|a}$) and ferromagnetic *c*-component (FM$_{\|c}$) of magnetic moment. No other IRs are compatible with the ferromagnetic moment along the *c* axis, which is observed in the bulk magnetization.

For quantitative analysis of the magnetic structure, the least-square refinement of the IR7 magnetic structure model was performed to the magnetic structure factors obtained at 2.5 K. To do this, we first performed structural refinement using the nuclear structure factors obtained at 7 K. Since we have the reflections only in the 0*kl* plane, the *x*-component of the fractional coordinate parameters are fixed to those obtained in the x-ray analysis. The result is given in Table III. In the magnetic structure refinement, we have fixed the atom fractional coordinate and isotropic displacement ($B_{iso}$) parameters to those obtained at 7.0 K to reduce the number of adjustable parameters. Consequently, we have two adjustable parameters, $m_a$ and $m_c$, in the magnetic structure refinement, which are the *a*- and *c*-components of the magnetic moment of Yb$^{3+}$. The refinement was performed using the temperature difference, $|F_{obs, 2.5 K}|^2 - |F_{obs, 7 K}|^2$, in order to minimize any effect due to structural ambiguity. The refined magnetic structure parameters and other refinement parameters are given in Table IV. The relation between the observed magnetic structure factor ($|F_{obs, 2.5 K}|^2 - |F_{obs, 7 K}|^2$) and the calculated magnetic structure factor ($|F_{mag, calc}|^2$) is plotted in Fig. 4 (a). The refined magnetic structure is visualized in Fig. 4 (b), indicating spontaneous ferromagnetic moment along the *c* axis with a finite antiferromagnetic component along the *a* axis. The magnetic moment of Yb$^{3+}$ was determined to be 0.78(3) μB, where the refined m$_a$ is 0.31(2) μB and the refined m$_c$ is 0.72(2) μB. The moment direction is 23 degrees away from the *c* axis, indicating that the non-collinear but coplanar AFM takes place at zero magnetic field in YbNiSn. It may be noted that the simulation of Scenario I [1] leads to very poor agreement with observed intensities.

**D. Proposed magnetic structure in the high-field phase ($H > H_{C1}$).**

Since a metamagnetic transition is observed at the critical field of $H_{C1}$ = 0.84 T along the *a* axis in bulk magnetization measurement at 2 K, we also performed the neutron diffraction experiment under external magnetic field. The YbNiSn single crystal was mounted to the cold head of the $^4$He refrigerator, and was set in an electromagnet with the highest magnetic field of 1 T. At zero field (Fig. 5 (a)), the scattering intensity at the 001 position shows a clear reduction near 5.7 K on heating. Under an applied field of 1 T along the *a* axis, the intensity shows no temperature dependence, as depicted in Fig. 5 (a). The $\theta$-$2\theta$ scans were performed for the selected 001 (Fig. 5 (b)) and 002 (Fig. 5 (c)) reflections both at 2.3 K ($< T_c$) and 7.0 K ($> T_c$). The 002 reflection shows a notable increase in scattering intensity, while no observable magnetic signal is seen for 001 reflection under the same condition. The results suggest the recovery of the reflection condition, 00*l* with *l* = 2*n*, in the high-field phase ($H$ = 1 T $> H_{C1}$ = 0.84 T).

The origin of metamagnetic transition could be associated with a spin-flip transition. Recalling the magnetic structure under zero magnetic field, shown in Fig. 4 (b), the *a*-components of Yb$^{3+}$ moments are antiparallel to each other. By applying the magnetic field along the *a*-axis, half of them can flip so that their *a*-components becomes parallel to reduce the Zeeman energy. Having this spin-flip transition in mind, we propose a putative magnetic structure under 1 T, as shown in Fig. 4 (c). Compared to the zero-field magnetic structure, the moments in the highlighted zone marked by dotted lines flip above $H_{C1}$. Accordingly, the magnetic space group of $Pn'm'a$, with a spontaneous magnetization along the *c* axis transforms into the magnetic space group of $Pnm'a'$, with a spontaneous magnetization now along the *a* axis. The magnetic structure after the spin-flip transition is still antiferromagnetic, as now the *c* components of the magnetic moments align antiparallelly. Both magnetic structures are non-collinear AFM but coplanar. This spin-flip transition naturally explains the recovery of the reflection condition, as the diagonal Yb moments in a unit cell now has a ferromagnetic *a*-axis component in the high-field phase.

Interestingly, we noticed that $Pn'm'a$ ($H < H_{C1}$) and $Pnm'a'$ ($H > H_{C1}$) commonly break the time reversal symmetry, whereas both of them still have the spatial inversion symmetry. Consequently, the piezomagnetic effect (PME) is theoretically allowed in both the phases, but in a distinct manner. For instance, a shear stress in the *ac* plane, $\sigma_{ac}$, provides induced magnetic moment along the *a* axis in $Pn'm'a$, whereas the *c*-axis component will be induced in $Pnm'a'$. Therefore, by applying external stress in a certain manner, it would be possible to switch the two magnetic phases. The switch could be rather easily detected by measuring the anomalous hall effect (AHE), since the direction of the bulk ferromagnetic moment is different for the two magnetic phases. Hence, we suggest that YbNiSn may be a rare candidate where the magnetic phase could be switched by the external stress, and such switch could be rather easily detected as the AHE switching.

## IV. CONCLUSION

The neutron and x-ray diffraction experiments were performed using Bridgman-grown single crystals of YnNiSn to determine its crystal and magnetic structures. At zero magnetic field, clear increase of magnetic intensity was observed below $T_c = 5.77(2)$ K. Magnetic structure analysis at the base temperature concludes that YbNiSn has a noncollinear canted antiferromagnetic structure with the magnetic space group $Pn'm'a$, rather than a simple collinear ferromagnetic structure proposed earlier. Under the external magnetic field of 1 T applied along the *a* axis, condition for the appearance of the magnetic reflections changes; it is highly likely that a spin-flip transition takes place at $H_{C1} = 0.84$ T, above which the magnetic space group changes to $Pnm'a'$. From the observed magnetic structures, we suggest that YbNiSn is a potential candidate to realize the magnetic phase switching, and accordingly AHE switching, by applying external stress.

**ACKNOWLEDGMENTS:**

The work at JRR-3 was supported by the General User Program of Neutron Science Laboratory, Institute for Solid State Physics, University of Tokyo. This work was supported by This work is partly supported by JSPS KAKENHI (Grants No. JP22H00101, JP19KK0069, JP19K21839, 19H01834, 19H05824, and 20K14395), and by "Crossover Alliance to Create the Future with People, Intelligence and Materials" from MEXT, Japan.

TABLE I. Results of the crystal structure refinement using the x-ray diffraction data measured at the room temperature. Obtained lattice constant, refined crystallographic parameters and reliability factors of YbNiSn are listed below. Occupancy is fixed to 1 for all the atoms. 2738 reflections with 383 unique reflections were used for the refinement. The independent refined parameters are 20.

**YbNiSn**

Orthorhombic *Pnma*, (No. 62)

$a = 6.965 (11)$ Å, $b = 4.416 (5)$ Å, $c = 7.618 (8)$ Å

$\alpha = 90°, \beta = 90°, \gamma = 90°$

| Atom | Site | x | y | z | $B_{iso}$ (Å$^2$) |
|---|---|---|---|---|---|
| Yb | 4c | 0.4894 (1) | 1/4 | 0.2969 (1) | 1.23 (2) |
| Ni | 4c | 0.2024 (3) | 1/4 | 0.5855 (3) | 1.41 (3) |
| Sn | 4c | 0.8075 (1) | 1/4 | 0.5863 (3) | 1.19 (2) |

$R = 4.14\%, wR = 10.4\%, GOF = 1.088$

TABLE II. The irreducible representation for the Yb 4(c) sites of the *Pnma* space group with the magnetic modulation vector $k = (0, 0, 1)$. Also shown are the magnetic structure category and magnetic space group corresponding to each irreducible representation.

| Irreducible representation | (x, ¼, z) | (-x + ½, ¾, z + ½) | (-x, ¾, -z) + (1, 0, 1) | (x + ½, ¼, -z + ½) | Magnetic category | Magnetic space group |
|---|---|---|---|---|---|---|
| IR1 | (0, $m_b$, 0) | (0, -$m_b$, 0) | (0, $m_b$, 0) | (0, -$m_b$, 0) | AFM$_{\|b}$ | *Pnma* |
| IR2 | (0, $m_b$, 0) | (0, $m_b$, 0) | (0, -$m_b$, 0) | (0, -$m_b$, 0) | AFM$_{\|b}$ | *Pn'ma* |
| IR3 | (0, $m_b$, 0) | (0, -$m_b$, 0) | (0, -$m_b$, 0) | (0, $m_b$, 0) | AFM$_{\|b}$ | *Pnma'* |
| IR4 | (0, $m_b$, 0) | (0, $m_b$, 0) | (0, $m_b$, 0) | (0, $m_b$, 0) | FM$_{\|b}$ | *Pn'ma'* |
| IR5 | ($m_a$, 0, $m_c$) | ($m_a$, 0, -$m_c$) | (-$m_a$, 0, -$m_c$) | (-$m_a$, 0, $m_c$) | AFM$_{\|ac}$ | *Pnm'a* |
| IR6 | ($m_a$, 0, $m_c$) | (-$m_a$, 0, $m_c$) | (-$m_a$, 0, -$m_c$) | ($m_a$, 0, -$m_c$) | AFM$_{\|ac}$ | *Pn'm'a'* |
| IR7 (H = 0 T) | ($m_a$, 0, $m_c$) | (-$m_a$, 0, $m_c$) | ($m_a$, 0, $m_c$) | (-$m_a$, 0, $m_c$) | AFM$_{\|a}$ + FM$_{\|c}$ | *Pn'm'a* |
| IR8 (H = 1 T) | ($m_a$, 0, $m_c$) | ($m_a$, 0, -$m_c$) | ($m_a$, 0, $m_c$) | ($m_a$, 0, -$m_c$) | FM$_{\|a}$ + AFM$_{\|c}$ | *Pnm'a'* |

Table III. Results of the crystal structure refinement using the nuclear intensity data obtained at 7.0 K. Unweighted single crystal reliability factor is defined as: $R_\mathrm{F} = 100 \times \sum_i [|y_{obs} - (\sum_i y_{cal}^2)^{\frac{1}{2}}|]/\sum_i y_{obs}$.

**YbNiSn**
Orthorhombic *Pnma*, (No. 62), $T = 7.0$ K (4G-GPTAS)
$a = $ n/a, $b = 4.3788$ Å, $c = 7.559$ Å
$\alpha = 90°, \beta = 90°, \gamma = 90°$

| Atom | Site | $x$ | $y$ | $z$ | $B_{iso}$ (Å$^2$) |
|---|---|---|---|---|---|
| Yb | 4c | 0.4894 | 1/4 | 0.2913 (15) | 1.506 (142) |
| Ni | 4c | 0.2024 | 1/4 | 0.5867 | 1.506 (142) |
| Sn | 4c | 0.8075 | 1/4 | 0.5929 | 1.506 (142) |

Number of nuclear reflections 13
Number of fitting parameter 2
$R_F = 6.17$ %

Table IV. Results of the magnetic structure refinement using the magnetic intensity data obtained as the temperature difference between 2.5 K and the paramagnetic temperature 7 K. The lattice constants shown below was the ones determined at 2.5 K using the 4G-GPTAS spectrometer. The crystal structure was assumed to be *Pnma*, and fractional-coordinate and isotropic-displacement parameters were fixed to those obtained in the crystal structure refinement, given in Table. III. The refined moment size and direction are summarized.

**YbNiSn**
Orthorhombic *Pnma*, (No. 62), $T = 2.5$ K (4G-GPTAS)
$a = $ n/a, $b = 4.3788$ Å, $c = 7.559$ Å
$\alpha = 90°, \beta = 90°, \gamma = 90°$

| Atom | Site | $x$ | $y$ | $z$ | $B_{iso}$ (Å$^2$) |
|---|---|---|---|---|---|
| Yb | 4c | 0.4894 | 1/4 | 0.2913 | 1.506 |

Number of magnetic reflections 9
Number of fitting parameter 2
Model: IR7 (*Pn'm'a*)

| Atom | Site | $m_a$ | $m_c$ |
|---|---|---|---|
| Yb | 4c | 0.31 (2) | 0.72 (2) |

$R_F = 2.79$ %

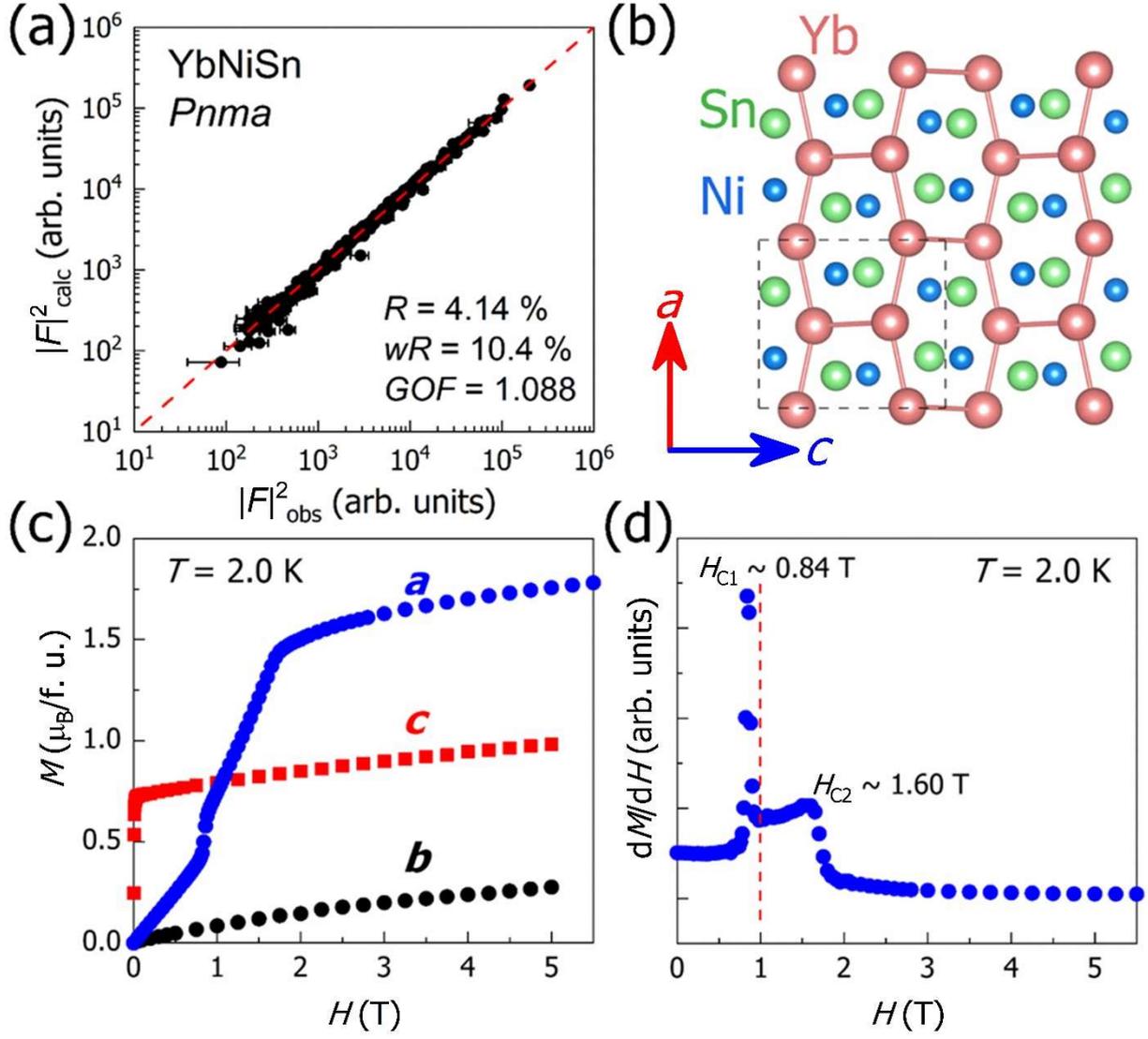

FIG. 1 (a) The comparison between the calculated $|F|^2_{calc}$ and observed $|F|^2_{obs}$ structure factors for x-ray. The observed ones were obtained using the YbNiSn single crystal the room temperature. (b) Schematic diagram of YbNiSn with orthorhombic TiNiSi-type structure obtained from the crystal structure refinement. The Yb atoms are represented by large red balls, while the Ni and Sn atoms are indicated by small blue and green balls, respectively. The crystal structure is made using VESTA [18]. (c) The isothermal $M$-$H$ curves up to 5 T of YbNiSn at 2.0 K for three crystallographic directions, indicating the existence of strongly anisotropic behavior. (d) The $dM/dH$ versus $H$ curve along $a$ direction at 2.0 K represents two independent critical fields at $H_{C1}$ = 0.84 T and $H_{C2}$ = 1.6 T respectively. The red dotted line represents the magnetic field of 1.0 T, where in-field neutron diffraction was performed.

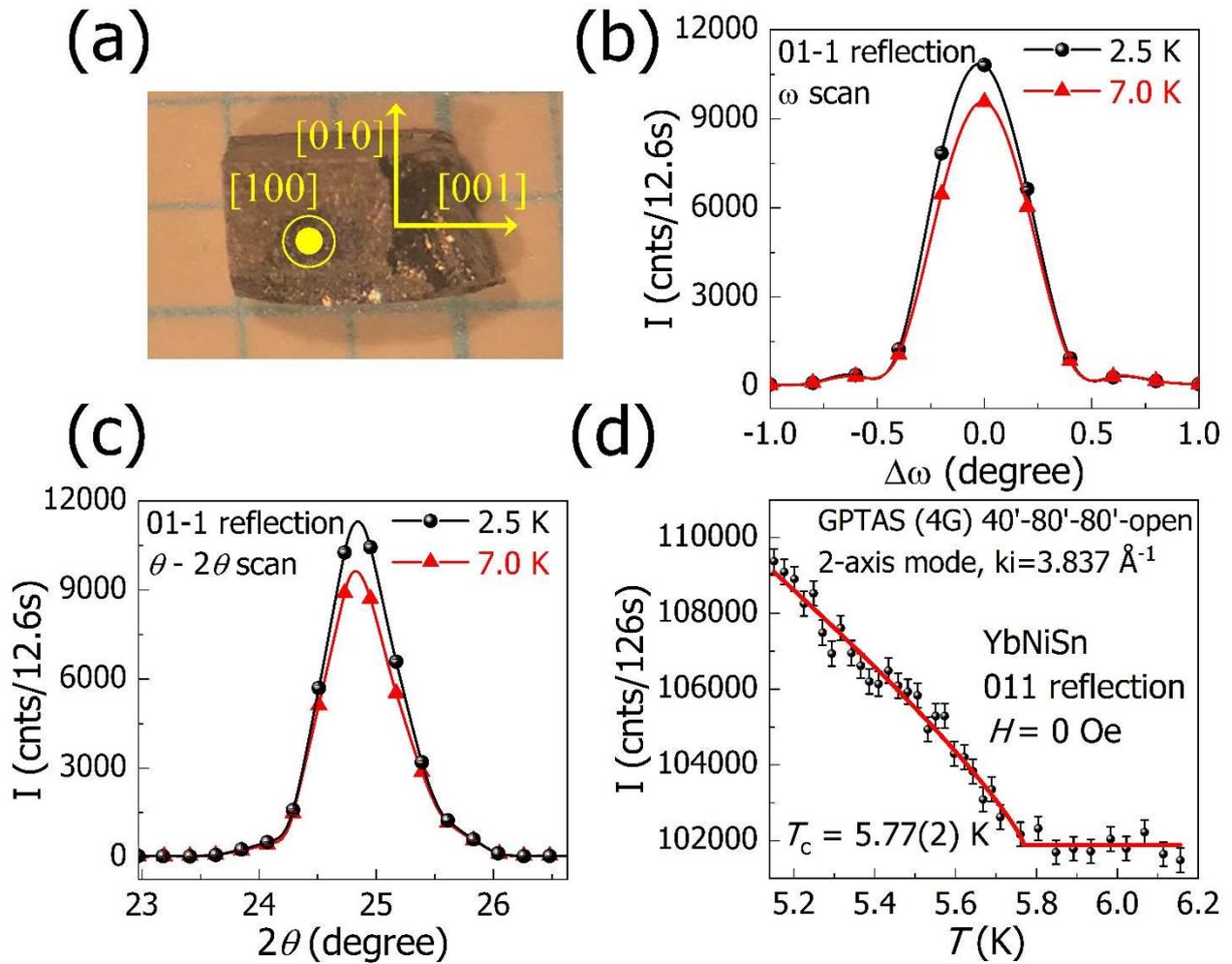

FIG. 2. (a) Photograph of YbNiSn single crystal on 1 mm graph paper. (b) The omega scans and (c) $\theta$-$2\theta$ scans were conducted at 2.5 K and 7.0 K to obtain the integrated intensity of the 01-1 reflection. (d) The temperature dependence of scattering intensity of the 011 reflection at zero-field. A magnetic transition is observed at $T_c$ = 5.77 (2) K.

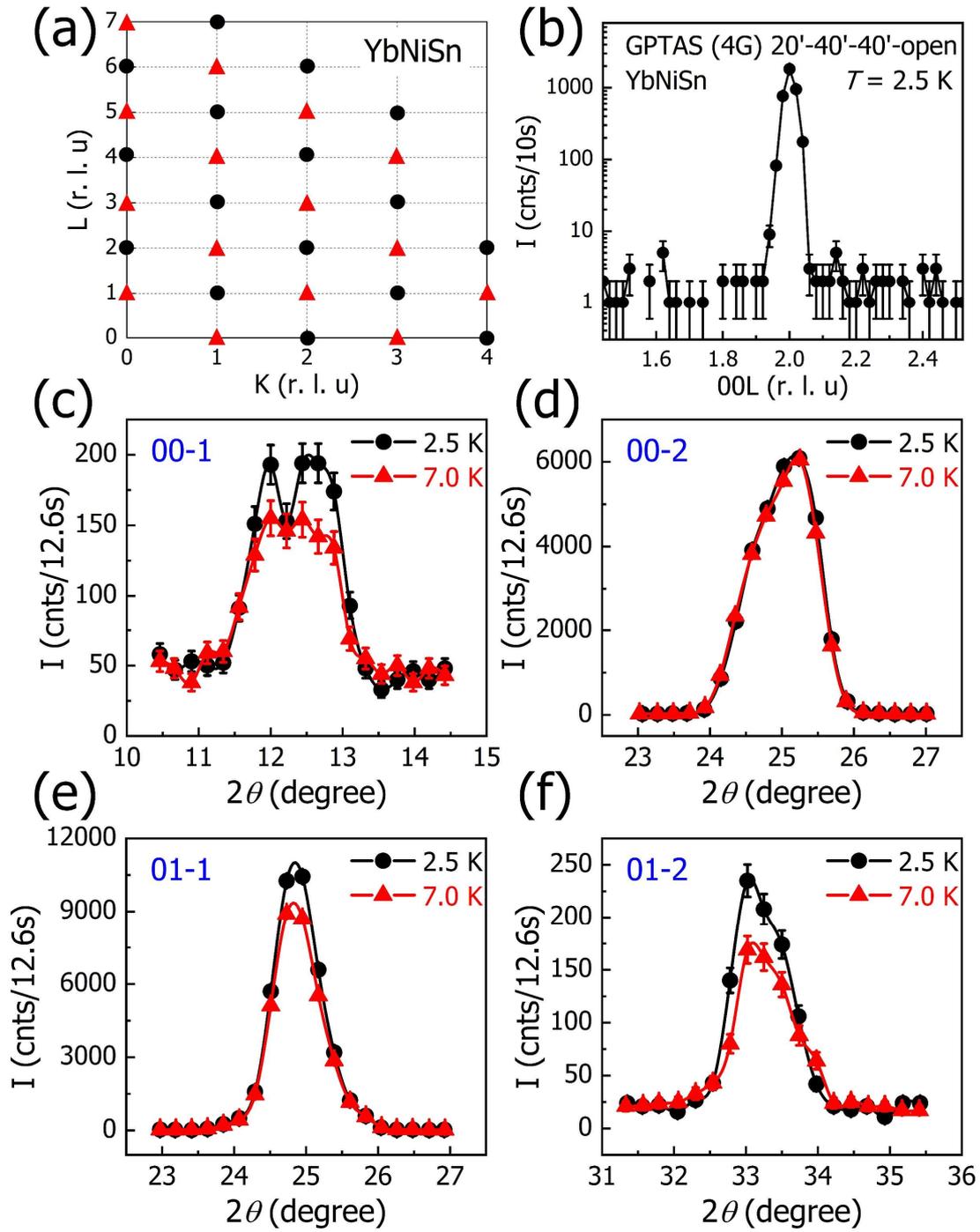

FIG. 3. (a) The schematic diagram depicts the $Q$ positions investigated in the $0kl$ scattering plane during the YbNiSn single crystal diffraction experiments. Black closed circles and red triangles represent the nuclear Bragg reflections and forbidden nuclear reflections, respectively. (b) The $l$-scan at 2.5 K along $00l$ direction around the 002 reflection. The $\theta$-$2\theta$ scans at 2.5 K for (c) 00-1, (d) 00-2, (e) 01-1, and (f) 01-2 reflections.

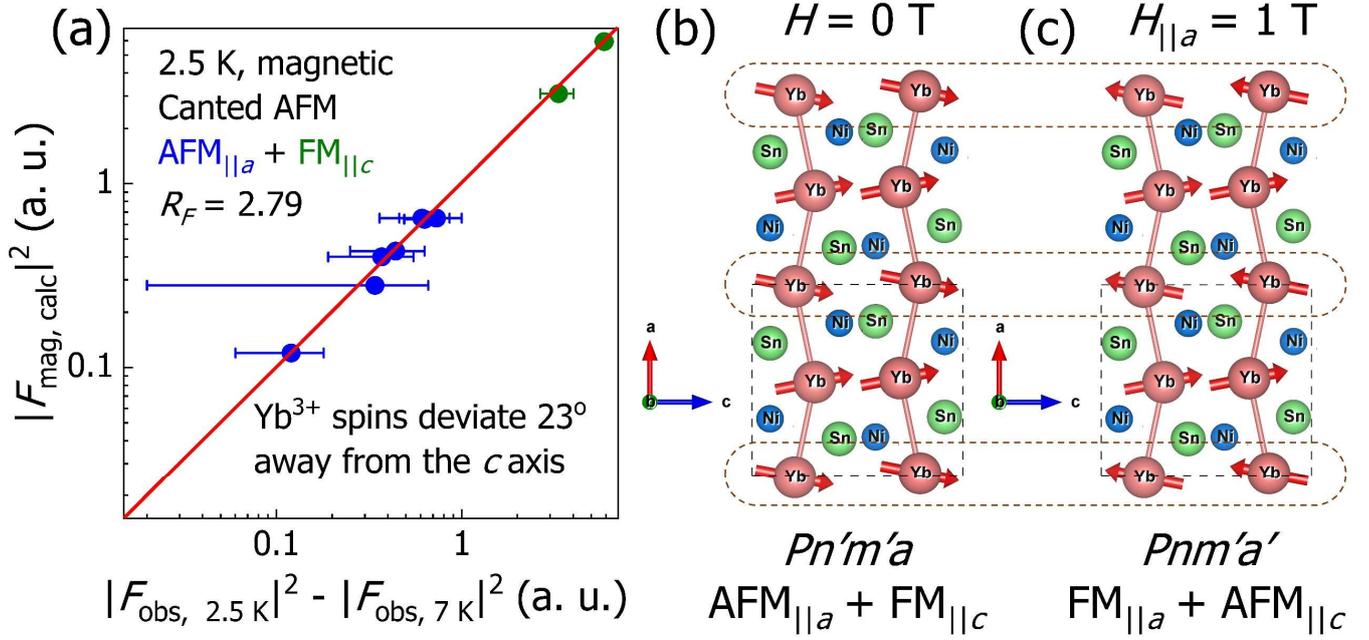

FIG. 4. (a) Squared magnetic structure factor $|F_{mag, calc}|^2$ calculated from the refined magnetic structure, versus observed magnetic structure factor obtained from the temperature difference $|F_{obs, 2.5 K}|^2 - |F_{obs, 7 K}|^2$. In (b), the refined magnetic structure is shown with $Yb^{3+}$ spins arranged in an antiferromagnetic configuration parallel to the $a$ axis and a ferromagnetic configuration parallel to the $c$ axis, with the spin direction deviating from the $c$ axis by 23 degrees. (c) Proposed magnetic structure for the high-field phase ($H > H_{C1}$). The area marked by dotted lines highlights the region where the spin-flip occurs. Both magnetic structures belong to non-collinear AFM but coplanar systems. The magnetic structure is generated using VESTA [18].

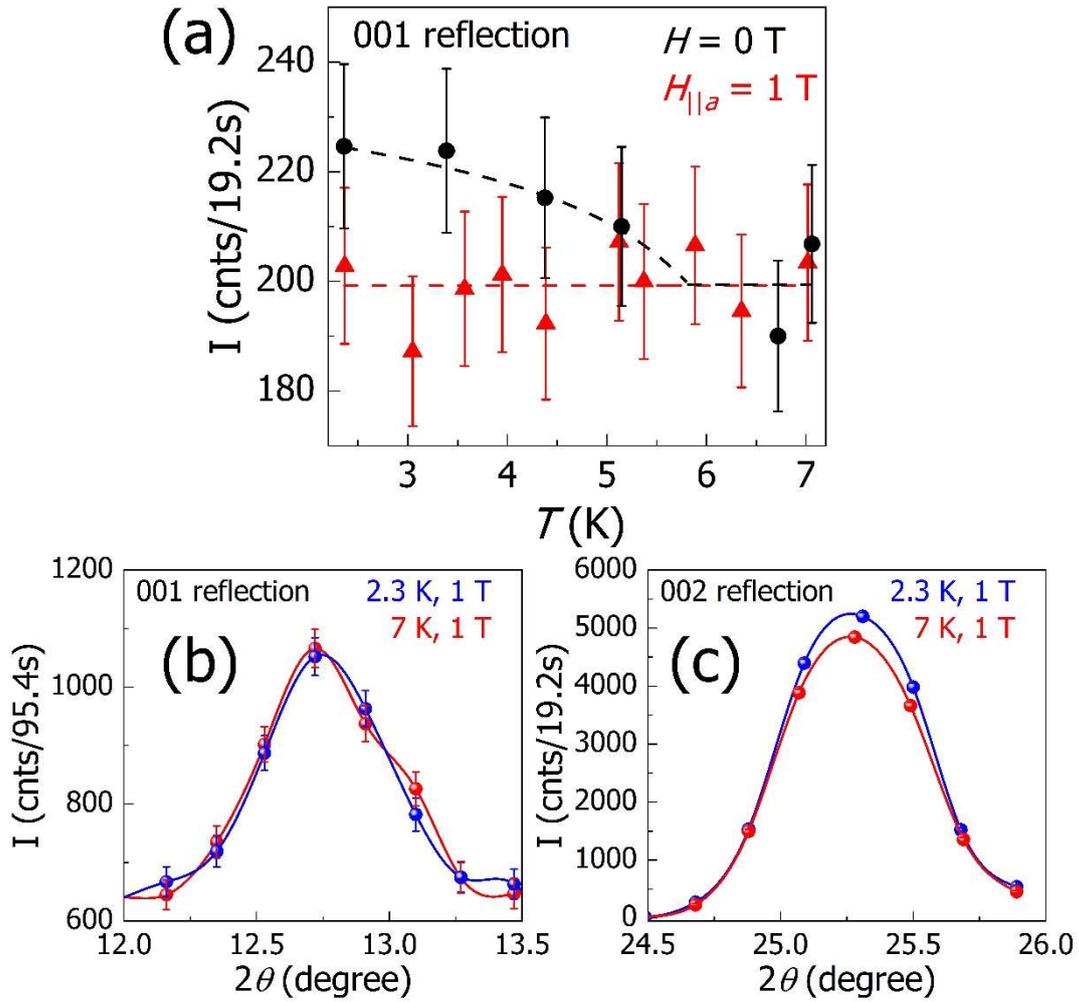

FIG. 5. (a) The temperature dependence of the scattering intensity of the 001 reflection measured under the external fields of 0 T and 1 T along the $a$ axis. The intensity increase is suppressed under 1 T. (b) and (c) The $\theta$-$2\theta$ scans for the 001 and 002 reflections measured at 2.3 K under 1 T and 7 K under 1 T. An increase in scattering intensity for the 002 reflection was observed, indicating a distinct reflection condition for zero-field and high-field ($H = 1$ T $> H_{C1}$) phases.